\documentstyle[12pt]{article}

\newcommand{\be}{\begin{equation}}
\newcommand{\ee}{\end{equation}}
\newcommand{\bea}{\begin{eqnarray}}
\newcommand{\eea}{\end{eqnarray}}

\begin{document}
\setlength{\baselineskip}{0.7cm}

\begin{titlepage}
\null
\begin{flushright}
hep-ph/0202196\\
UT-02-10\\
February, 2002
\end{flushright}
\vskip 1cm
\begin{center}
{\Large\bf 
Decoupling Solution to SUSY Flavor Problem \\
via Extra Dimensions 
}

\lineskip .75em
\vskip 1.5cm

\normalsize

{\large Naoyuki Haba$^{a}$}
and {\large Nobuhito Maru$^{b}$}

\vspace{1cm}

{\it $^{a}$Faculty of Engneering, Mie University, Tsu, 
Mie, 514-8507, JAPAN} \\
{\it $^{b}$Department of Physics, University of Tokyo, 
Tokyo 113-0033, JAPAN} \\


%
\vspace{18mm}

{\bf Abstract}\\[5mm]
{\parbox{13cm}{\hspace{5mm}
%
We discuss the decoupling solution to SUSY flavor problem 
in the fat brane scenario. 
We present a simple model to yield 
the decoupling sfermion spectrum 
in a five dimensional theory. 
Sfermion masses are generated by the overlap between 
the wave functions of the matter fields and 
the chiral superfields on the SUSY breaking brane. 
Two explicit examples of the spectrum are given. 
}}

\end{center}

\end{titlepage}

In building models with supersymmetry (SUSY), 
we must take into account that 
the sfermion mass spectrum of the first and 
the second generations is severely constrained from 
the flavor changing neutral current (FCNC) processes 
such as $K^0-\bar{K}^0$ mixing etc. 
Mainly, three approaches to address this problem 
(SUSY flavor problem) have been discussed in the literature; 
(1)~the degeneracy \cite{dege}, 
(2)~the alignment \cite{NS} and 
(3)~the decoupling \cite{deco}.

In this letter, 
we consider the decoupling solution to SUSY flavor problem 
in the context of higher dimensional theories. 
The basic idea is very simple. 
In extra dimensions, 
it is well known that if the matter wave functions are localized 
at different points in extra dimensions, 
Yukawa hierarchy can be obtained by the suppression factor of 
the overlap of wave functions \cite{AS}. 
Since the fermion mass hierarchy is $m_1 < m_2 < m_3$ 
where $m_i$ is a fermion mass of the i-th generation, 
the matter of the third generation is localized close to 
the Higgs fields and the first generation is localized 
most distant from the Higgs fields. 
Introducing SUSY breaking brane in which 
the chiral superfield 
with nonvanishing F-term is localized correlates 
the sfermion masses with the fermion masses. 
If the SUSY breaking brane is put 
 close to the first generation matter fields, 
the sfermion mass hierarchy is inverted, 
$\tilde{m}_1 > \tilde{m}_2 > \tilde{m}_3$ 
where $\tilde{m}_i$ is the sfermion mass of the i-th generation. 
Therefore, we expect that 
the decoupling solution can be a natural solution. 
Namely, the sfermion masses of the first and 
the second generation is the order of 10 TeV and 
the sfermion mass of the third generation is the order of 100 GeV 
for naturalness.

Let us discuss the model in detail. 
We consider an ${\cal N}=1$ supersymmetric theory in five dimensions. 
We introduce two 3-branes at $y=0$ and $y = L$, 
where $y$ denotes the fifth coordinate 
in five dimensional space-time. 
The gauge supermultiplets of the Standard Model (SM) 
gauge groups lives in the bulk and 
its zero mode wave functions are flat in the fifth dimension. 
The matter fields also lives in the bulk and 
its zero mode wave functions are assumed 
to be Gaussian.\footnote{For 
readers interested in the localization mechanism 
of the chiral superfields, see Appendix of Ref.~\cite{KT}.} 
Higgs doublets are assumed to be localized 
on the brane at $y=0$, 
we refer to this brane as ``H-brane". 
Further, extra chiral superfields $X, \Phi'$ 
and $\bar{\Phi}'$ 
localized on the brane at $y=L$ are introduced. 
$X$ is a chiral superfield with nonvanishing F-term 
(i.e. $X=\theta^2F$). 
$\Phi'$ and $\bar{\Phi}'$ are vector-like superfields 
with a mass $M$. 
A pair of vector-like superfields are introduced 
for each matter chiral superfields, 
namely $Q', \bar{Q}'$ for $Q$, and $U', \bar{U}'$ for $\bar{U}$ 
and $L', \bar{L}'$ for $L$ etc. 
We refer to the brane at $y=L$ as ``SUSY breaking brane".

Naive expectation is that 
the gaugino mass is generated from the term 
\bea
\delta(y-L) \int d^2\theta \frac{X(x)}{M_*^2} 
W^\alpha(x,y) W_\alpha(x,y) 
\to M_{\lambda} = \frac{F}{M_*^2L_c}, 
\eea
where $x$ is a coordinate of the four dimensional space-time, 
$W_\alpha$ is the field strength tensor 
superfield living in the bulk, 
$L_c$ is the width of the thick wall 
which should be considered as the compactification length 
in our framework and 
$M_*$ is the Planck scale in five dimensions. 
Note that the gaugino masses receive only 
the volume suppression factor. 
The soft breaking mass of the $i$-th 
 generation chiral superfield $\Phi_i$ 
 is naively generated from the term 
\bea
&&\delta(y-L) \int d^4\theta \frac{X^\dag (x)X(x)}{M_*^3} 
\Phi_i^\dag (x,y) \Phi_j (x,y) \nonumber \\
&&\to \tilde{m}^2_{ij} = 
\frac{|F|^2}{M_*^3L_c}~
{\rm exp}[-M_*^2(L-y_i)^2-M_*^2(L-y_j)^2], 
\eea
where the form of the zero mode wave function of 
the matter fields is assumed as 
$\Phi_i^{(0)} \sim {\rm exp}[-M_*^2(y-y_i)^2]$. 
The sfermion masses receives not only the volume suppression 
but also the exponential suppression, 
therefore are negligibly small compared to the gaugino masses. 
This spectrum is similar to the gaugino mediation 
scenario \cite{gMSB}. 
Although the spectrum of the gaugino mediation 
is phenomenologically interesting, 
this is not the subject in this paper.

As is clear from the above argument, 
the modification is needed to obtain the spectrum of 
the decoupling solution. 
The wayout is to replace $M_*$ with some scale $M<M_*$ 
such that the enhancement by $M$ compensates 
for the exponential suppression. 
We consider here the following superpotential 
on the SUSY breaking brane 
\bea
W &=& \delta(y-L) \int dy
[\lambda X(x) \Phi_i(x,y) \bar{\Phi}'(x) + M \Phi'(x) 
\bar{\Phi}'(x)], \\
&=& \lambda \epsilon_i X(x) \Phi_i(x) \bar{\Phi}'(x) 
+ M \Phi'(x) \bar{\Phi}'(x), 
\eea
where $\lambda$ is a dimensionless constant of order unity. 
The second expression is obtained by integrating out the fifth 
dimensional degree of freedom. 
$\epsilon_i$ is a suppression factor coming from the zero mode 
wave function of i-th generation. 
At the scale below $M$, 
we can integrate out the massive fields $\Phi', \bar{\Phi}'$. 
Then, the effective superpotential of Eq.(4) vanishes, and 
the effective K\"ahler potential receives the correction 
at tree level such as 
\be
\label{Kahler}
\delta K = \frac{\epsilon_i^2}{M^2} X^\dag X \Phi_i^\dag \Phi_i. 
\ee
As mentioned above, 
it is necessary to introduce a pair of 
vector-like fields for each matter chiral superfield 
to obtain the above K\"ahler potential. 
Otherwise, 
the K\"ahler potential with the flavor mixing will 
arise in general. 
We does not consider this possibility. 
Sfermion masses coming from Eq.~(\ref{Kahler}) becomes 
\bea
\tilde{m}^2_i = \epsilon_i^2 \frac{|F|^2}{M^2}. 
\eea
This result has the desirable features that 
the suppression factor is not only replaced with $M<M_*$, 
but also proportional to $\epsilon \simeq ({\rm Yukawa})^{-1}$. 
Note that this argument holds in the case $F < M^2$. 
Also, we assumed here that the overall sign of the K\"ahler potential 
is positive.

In our scenario, 
since the information of the location where the matter fields are localized 
is necessary to derive the sfermion mass spectum, 
we briefly discuss Yukawa hierarchy. 
We consider the up-type Yukawa coupling for example 
\bea
W = \delta(y) \int dy Q_i(x,y) \bar{U}_j(x,y) H(x), 
\eea
where the order one coefficient is implicit. 
Integrating out the fifth dimensional degree of freedom, 
we obtain the effective Yukawa coupling in four dimensions 
at the compactification scale $L_c^{-1}$ as 
\bea
\label{effyukawa}
y_{{\rm eff}} \simeq {\rm exp}[-M_*^2(y_{Q_i}^2 - y_{\bar{U}_j}^2)]. 
\eea
In order to realize the Yukawa hierarchy 
\bea
\label{yukawa1}
&&y_t \sim {\cal O}(1),~y_c \sim {\cal O}(10^{-2}),
~y_u \sim {\cal O}(10^{-5}), \\
&&y_b \sim {\cal O}(10^{-2}),~y_s \sim {\cal O}(10^{-4}),
\label{yukawa2}
~y_d \sim {\cal O}(10^{-5}), 
\eea
the location of the matter fields are determined,\footnote{To be correct, 
the effective Yukawa couplings (\ref{effyukawa}) 
have to be evolved down to the weak scale 
by the renormalization group equation (RGE) 
and matched to Eqs.~(\ref{yukawa1}), (\ref{yukawa2}) 
after diagonalizing Yukawa matrix.
We simply neglect this RGE effects and the mixing angles.} 
for instance, 
%
\bea
&&y_{H} \simeq y_{\bar{H}} \simeq y_{Q_3} \simeq y_{U_3} 
\simeq 0,~|y_{Q_2}| \simeq |y_{U_2}| \simeq \sqrt{{\rm ln}10} M_*^{-1},
~|y_{D_3}| \simeq \sqrt{2{\rm ln}10} M_*^{-1}, \\
&&|y_{Q_1}| \simeq |y_{U_1}| \simeq |y_{D_1}| 
\simeq \sqrt{\frac{5}{2}{\rm ln}10} M_*^{-1},~
y_{D_2} \simeq \sqrt{3{\rm ln}10} M_*^{-1}. 
\eea

Now, we turn to the sfermion mass spectrum. 
The decoupling solution requires that 
the masses of the first and the second generations 
should be heavier than ${\cal O}$(10) TeV and 
that of the third generation be around ${\cal O}$(100) GeV 
for naturalness. 
We consider two cases of the decoupling constraints
 according to Eqs.(11) and (12). 
The first case is 
 $\tilde{m}_{Q_2} = \tilde{m}_{U_2} 
 \simeq 10$ TeV (solution 1), 
 and the second one is $\tilde{m}_{D_2} \simeq 10$ TeV (solution 2).

Let us consider the solution 1 at first. 
This case imposes  
\bea
\label{decoupling}
10^2 \simeq \epsilon_2/\epsilon_3 
\simeq {\rm exp}[(M_* L)^2-(M_* L - \sqrt{{\rm ln}10})^2], 
\eea
which leads to 
\bea
\label{L}
M_* L \simeq \frac{3}{2}\sqrt{{\rm ln}10}. 
\eea
This implies that SUSY breaking brane is located 
between the first and 
the second gengerations. 
This is the interesting feature of our model. 
In the conventional SUSY breaking models in extra dimensions, 
the visible brane and the SUSY breaking brane are separated 
in the extra dimensional spaces and the dangerous 
flavor violating sfermion masses are suppressed by the locality. 
In our model, 
SUSY breaking brane are not separate from the visible brane 
(more correctly, the visible wall) but is predicted to be located 
within the visible wall to yield the decoupling sfermion mass spectrum. 

Requiring that the gaugino mass should be around 100 GeV, 
\bea
M_{\lambda} \simeq \frac{F}{M_*}\frac{1}{M_* L_c} 
\simeq 100~{\rm GeV}, 
\eea
we obtain 
\bea
\label{F}
F \simeq (M_* L_c) \times 10^2 M_*. 
\eea
%
Since $\tilde{m}_{Q_2} = \tilde{m}_{U_2} \simeq$ 10 TeV, 
\bea
\tilde{m}_{Q_2} = \tilde{m}_{U_2} & \simeq & \frac{F}{M}
~{\rm exp}[-(M_* L - \sqrt{{\rm ln}10})^2], \\
&\simeq& \frac{10^2(M_* L_c)(M_* L)}{ML}~
{\rm exp}[-(\frac{3}{2} \sqrt{{\rm ln}10} - \sqrt{{\rm ln}10})^2] 
\simeq 10~{\rm TeV}, \\
&\Leftrightarrow& 
\label{M}
ML \simeq (M_* L_c) \times 10^{-2} \times \frac{3}{2}
\sqrt{{\rm ln}10}~{\rm exp}[-{\rm ln}10/4]. 
\eea
Using Eqs.~(\ref{L}, \ref{F}, \ref{M}), 
we obtain the sfermion masses for other generations 
\bea
&&\tilde{m}_{Q_1} = \tilde{m}_{U_1} =\tilde{m}_{D_1} 
\simeq \frac{F}{M}~{\rm exp}
[-(M_*L-\sqrt{5{\rm ln}10/2})^2] \simeq 18~{\rm TeV}, \\
&&\tilde{m}_{Q_3} = \tilde{m}_{U_3} \simeq \frac{F}{M}~{\rm exp}
[-9{\rm ln}10/4] \simeq 100~{\rm GeV}, \\
&&\tilde{m}_{D_2} \simeq \frac{F}{M}~{\rm exp}
[-(3\sqrt{{\rm ln10}}/2 - \sqrt{3{\rm ln10}})^2] 
\simeq 15.7~{\rm TeV}, \\
&&\tilde{m}_{D_3} \simeq \frac{F}{M}~{\rm exp}
[-(3\sqrt{{\rm ln10}}/2 - \sqrt{2{\rm ln10}})^2] 
\simeq 17.5~{\rm TeV}. 
\eea
We note that these sfermion masses are generated at 
the compactification scale $L_c^{-1}$.

Let us comment on scales in our model. 
Assuming $M_* \simeq 10^{18}{\rm GeV}$ and 
$M_* L_c \simeq 10$, 
\bea
F \simeq 10^{21}{\rm GeV^2},
~M \simeq 5.6 \times 10^{16}{\rm GeV},
~L_c^{-1} \simeq 10^{17}{\rm GeV}
\eea
are obtained. 
As mentioned above, 
$F < M^2$ is satisfied. 
It is also interesting that the mass of the additional 
vector-like superfields is close to the compactification 
scale.\footnote{Since the mass of the vector-like fields $M$ 
is a parameter in four dimensional theory, 
it has nothing to do with 
the fundamental Planck scale $M_*$.}

Next we will show the case of the solution 2. 
Instead of the condition (\ref{decoupling}), 
we can impose  
\bea
10^2 \simeq \epsilon_2/\epsilon_3 \simeq 
{\rm exp}[(M_* L)^2 - (M_* L - \sqrt{3{\rm ln10}})^2], 
\eea
which leads to 
\bea
M_*L \simeq \frac{5}{6}\sqrt{3{\rm ln10}}. 
\eea
{}From the gaugino mass constraint, we obtain 
\bea
F \simeq (M_*L_c) \times 10^2M_*. 
\eea
On the other hand, 
 $\tilde{m}_{D_2} \simeq $10 TeV leads to 
\bea
ML \simeq (M_* L_c) (ML) \times 10^{-2}{\rm exp}[-{\rm ln}10/12]. 
\eea
Other sfermion masses 
 at the compactification scale 
 can be estimated 
 as the same anaylses of the previous case as, 
\bea
&&\tilde{m}_{Q_3} = \tilde{m}_{U_3} \simeq 100~{\rm GeV},\;\;
~\tilde{m}_{D_3} \simeq 12.1~{\rm TeV}, \\
&&\tilde{m}_{Q_2} = \tilde{m}_{U_2} \simeq 7.7~{\rm TeV},\;\;
~\tilde{m}_{Q_1} = \tilde{m}_{U_1} = \tilde{m}_{D_1} \simeq 11.6~{\rm TeV}. 
\eea
As for scales, it is almost the same as the solution 1, 
\bea
F \simeq 10^{21}{\rm GeV^2},\;
~M \simeq 8.25 \times 10^{16}{\rm GeV},\;
~L_c^{-1} \simeq 10^{17}{\rm GeV}. 
\eea

We give some comments here. 
First, 
in our scenario, $\tilde{m}_{D_3}$ is the order of 
${\cal O}$(10) TeV, which is somewhat large 
from the viewpoint of 
the decoupling solution. 
This means that the large tan$\beta$ case is preferable. 
Second, 
it is known that the decoupling scenario generically suffers from 
a problem that the third generation sfermion mass squareds 
are driven to be negative through the two-loop RGE effects of 
the heavy first-two generation sfermion masses \cite{negative}. 
%
%
Since the sfermion masses we obtained are generated at 
the compactification scale, the third generation sfermion mass 
squareds might be negative at low energy. 
To avoid this, 
we have to add the extra fields with negative SUSY breaking 
mass squareds as discussed in Ref.~\cite{hkn}. 
We do not discuss this point in detail in this paper.

In summary, 
we have proposed the decoupling scenario in the fat brane approach. 
In this approach, Yukawa hierarchy is determined by the overlap 
of wave functions of the matter fields 
localized at the different points 
in extra dimensions. 
The lighter matter fileds are localized closer to 
the point where Higgs fields are localized. 
We introduced the SUSY breaking brane in which the chiral superfield 
with nonzero F-term is localized and discussed 
whether the spectrum of the decoupling solution 
is possible or not, 
namely, the firsr-two generation sfermion masses are the order 
of 10 TeV, and the third generation sfermion masses are of order 100 GeV. 
Naively, sfermion masses generated from the K\"ahler potential 
suppressed by the fundamental scale are negligibly small 
compared to the gaugino masses since sfermion masses receive 
the additional exponential suppression 
by the overlap between the wave functions of 
the matter firlds and 
of the chiral superfield on the SUSY breaking brane. 
In order to obtain the spectrum of the decoupling scenario, 
we have introduced the extra vector-like superfields 
and Yukawa interaction between the vector-like superfields, 
MSSM superfields and the chiral superfield with nonzero F-term. 
Integrating out the vector-like fields leads to 
the K\"ahler potential proportional to the suppression factor 
with the inverse of Yukawa hiearchy and suppressed by 
the mass of the vector-like fields, which is smaller than 
the fundamental scale. 
This becomes the dominant source of the sfermion masses. 
Two explicit examples of the spectrum have been shown in this paper. 
It has turned out that 
the large tan$\beta$ case is preferable because $\tilde{m}_{D_3}$ 
is somewhat large. 
It is interesting that SUSY breaking brane is predicted to 
be localized within the thick visible wall 
unlike the conventional SUSY breaking scenario in extra dimensions.

Finally, we have touched on the negative sfermion mass 
squareds problem. 
Since the sfermion masses we have obtained is generated 
at the compactification scale, 
a detailed RGE analysis is necessary to see 
whether the third generation sfermion mass squareds 
are indeed positive at low energy. 
We leave this subject for future work. 

\vspace{1cm}

\begin{center}
{\bf Acknowledgments}
\end{center}
N.H. is supported by the Grant-in-Aid for Scientific Research, 
Ministry of Education, Science and Culture, Japan (No.12740146) and 
N.M. is supported 
by the Japan Society for the Promotion of Science 
for Young Scientists (No.08557).

\vspace{1cm}


\begin{thebibliography}{99} 
 \bibitem{dege}
   See, for example, J.S.~Hagelin, S.~Kelly and T.~Tanaka, 
    {\em Nucl. Phys.} {\bf B415} (1994) 293; 
   F.~Gabbiani, E.~Gabrielli, A.~Masiero and L.~Silvestrini, 
    {\em Nucl. Phys.} {\bf B477} (1996) 321 [hep-ph/9604387]. 
 \bibitem{NS}
   Y.~Nir and N.~Seiberg, 
     {\em Phys. Lett.} {\bf B309} (1993) 337 [hep-ph/9304307]. 
 \bibitem{deco}
   M.~Dine, A.~Kagan and S.~Samuel, 
     {\rm Phys. Lett.} {\bf B243} (1990) 250; 
   S.~Dimopoulos and G.F.~Giudice, 
     {\em Phys. Lett.} {\bf B357} (1995) 573 [hep-ph/9507282]; 
   A.~Pomarol and D.~Tommasini, 
     {\em Nucl. Phys.} {\bf B466} (1996) 3 [hep-ph/9507462]; 
   A.G.~Cohen, D.B.~Kaplan and A.E.~Nelson, 
     {\em Phys. Lett.} {\bf B388} (1996) 588 [hep-ph/9607394]. 
 \bibitem{AS}
   N.~Arkani-Hamed and M.~Schmaltz, 
    {\em Phys. Rev.} {\bf D61} (2000) 033005 [hep-ph/9903417]; 
   E.A.~Mirabelli and M.~Schmaltz, 
   {\em Phys. Rev.} {\bf D61} (2000) 113011 [hep-ph/9912265]. 
 \bibitem{KT}
   D.E.~Kaplan and T.M.P.~Tait, 
    {\em JHEP} {\bf 0006} (2000) 020 [hep-ph/0004200]. 
 \bibitem{gMSB}
   D.E.~Kaplan, G.D.~Kribs and M.~Schmaltz, 
    {\em Phys.~Rev.} {\bf D62} (2000) 035010 [hep-ph/9911293]; 
   Z.~Chacko, M.A.~Luty, A.E.~Nelson and E.~Ponton, 
    {\em JHEP} {\bf 0001} (2000) 003 [hep-ph/9911323]. 
 \bibitem{negative}
   N.~Arkani-Hamed and H.~Murayama, 
    {\em Phys. Rev.} {\bf D56} (1997) 6733 [hep-ph/9703259]; 
   K.~Agashe and M.~Graesser, 
    {\em Phys. Rev.} {\bf D59} (1999) 015007 [hep-ph/9801446]. 
 \bibitem{hkn}
   J.~Hisano, K.~Kurosawa and Y.~Nomura, 
    {\em Nucl. Phys.} {\bf B584} (2000) 3 [hep-ph/0002286]. 
\end{thebibliography}
\end{document}